\documentclass[submission,copyright,creativecommons]{eptcs}
\providecommand{\event}{ICLP 2022} 
\usepackage{breakurl}             
\usepackage{underscore}           
\usepackage{xspace}
\usepackage{amsmath}

\newcommand{\clingo}{{\sc clingo}\xspace}
\newcommand{\anthem}{{\sc anthem}\xspace}
\newcommand{\vampire}{{\sc vampire}\xspace}
\newcommand{\ruleo}{\;\hbox{:-}\;}

\begin{document}

\title{Tools and Methodologies for Verifying Answer Set Programs}

\author{Zach Hansen
\institute{University of Nebraska Omaha\\Omaha, Nebraska}
\email{zachhansen@unomaha.edu}
}

\def\titlerunning{Tools and Methodologies for Verifying Answer Set Programs}
\def\authorrunning{Hansen}

\maketitle










\section{Introduction}
Answer Set Programming (ASP) is a powerful declarative programming paradigm commonly used for solving challenging search and optimization problems~\cite{Marek1999,Niemela1999}. The modeling languages of ASP are supported by sophisticated solving algorithms (solvers) that make the solution search efficient while enabling the programmer to model the problem at a high level of abstraction~\cite{Janhunen2016}. As an approach to Knowledge Representation and Reasoning, ASP benefits from its simplicity, conciseness and rigorously defined semantics. These characteristics make ASP a straightforward way to develop formally verifiable programs. In the context of artificial intelligence (AI), the clarity of ASP programs lends itself to the construction of explainable, trustworthy AI. In support of these goals, my research is concerned with extending the theory and tools supporting the verification of ASP progams.

The formal verification of ASP programs presents several challenges that differentiate it from the verification of procedural imperative software. First and foremost, ASP programs lack modularity; traditionally, the meaning of a portion of code cannot be determined in isolation from the remainder of the program. Second, since the program semantics are traditionally defined through grounding, it is difficult to reason formally about the correctness of a program without referring to a specific problem instance. Finally, the well-established tools of proof for reasoning about procedural imperative languages (such as loop invariants) are not directly applicable to ASP programs. Instead, we must explore new verification strategies, both manual and automated, to provide a high level of assurance about our programs' behavior.

\section{Background}
Approaches to defining semantics for logic programs can be roughly divided into two broad categories: fixpoint formalisms, and translational formalisms~\cite{feleli11a}. 
The original definition of a stable model from 1988 is an example of the fixpoint approach; it defines the behavior of basic logic programs with negation in terms of reducts~\cite{gel88}. Advances in grounding and solving technology have made the implementation of these semantics efficient, however, reasoning about the correctness of programs in terms of fixpoints can be challenging. 
Part of the challenge is the inseperability of problem class and problem instance. 
Another challenge faced by fixpoint approaches is the definition of advanced language constructs, such as arbitrary choice rules, conditional literals, and aggregates. These are useful language features, but their behavior in terms of the grounding and solving process is typically described informally~\cite{harliffang14}. 
Thus, most of my research so far has focused on developing alternative translational semantics for advanced ASP language constructs.

One example of a translational approach is {\em program completion}~\cite{Clark1978,halira17a}. Programs meeting the syntactic requirement of {\em tightness}\footnote{A program is {\em tight} if it has an acyclic dependency graph, for details see~\cite{cab20b}.} can be converted into a first-order sentence whose models are in one-to-one correspondence with the answer sets of the program. A related approach is the SM operator~\cite{feleli11a}. Much like completion, applying the SM operator to a logic program first requires a syntactic transformation into a first-order sentence. The SM operator then transforms this sentence into a second-order one and uses predicate quantification to minimize belief in certain {\em intensional} predicates. When all the predicates occurring in the logic program are treated as intensional, then the models of this second-order sentence correspond to the program's answer sets. These translational approaches have the advantage of avoiding grounding entirely. 

\anthem is a software system that transforms a non-ground ASP program into an equivalent first-order theory via program completion~\cite{falilusc20a,liflusch}. Given an ASP program and a formal specification written in first-order logic, \anthem translates the program into typed first-order theories and then uses the theorem prover \vampire~\cite{kovvor13a} to verify the adherence of the translation to the specification. This provides a way to automatically verify the correctness of tight logic programs. We wish to extend the capabilities of \anthem to include a broader class of logic programs, such as those containing aggregates and conditional literals.
The semantics implemented by the answer set solver \clingo of these language constructs are given by a translation~\cite{harliffang14} to infinitary propositional logic~\cite{tru12}.

\section{Current Status}
Since joining the University of Nebraska Omaha in Fall 2020, I have been engaged in several research projects under the umbrella topic of {\em formal verification of ASP programs}. I have worked with my advisor (Dr. Yuliya Lierler) and Dr. Jorge Fandinno to define the semantics of non-basic ASP language constructs such as aggregates and conditional literals in a manner convenient for proving results about their behavior. Additionally, we have extended the modular proof methodology developed by Cabalar, Fandinno, and Lierler (2020) to programs with aggregates occurring in constraints.

\subsection{Many-sorted semantics for aggregates (\cite{fan22})}
\label{sec:agg}
A project I have recently worked on with Dr. Lierler and Dr. Fandinno provided an alternative characterization of aggregate semantics, which are typically defined through a translation to infinitary propositional logic. Conversely, our approach avoids referring to grounding by applying a many-sorted generalization of the SM operator to a set of many-sorted first-order formulas representing a logic program. Aggregates are defined as functions on sets of tuples, whose members are restricted to those tuples satisfying the list of conditions present in the associated aggregate. To ensure this behavior, we add second-order axioms to the program that fix the behavior of sets and aggregate function symbols. We proved the equivalence of our semantics to the semantics implemented by \clingo for programs that do not contain positive recursion through aggregates. Furthermore, for tight programs with finite aggregates, we can replace the second-order axiomatization with a first-order one. This results in a fully first-order characterization of the behavior of these programs.

This contribution helps to address one of the fundamental issues identified in the Introduction; namely, it decouples the argument of program correctness from specific problem instances for programs with aggregates. Since aggregates are such useful and common constructs, this  substantially broadens the class of ASP programs with formally defined non-ground semantics. We also envision this work as part of the foundation required to automatically verify the correctness of programs with aggregates. The long-term goal of this line of research is to extend the capabilities of \anthem accordingly (Section~\ref{sec:anthem}).

\subsection{Many-sorted semantics for conditional literals}
\label{sec:ms:cl}
In a similar vein as the project from Section~\ref{sec:agg}, I developed a many-sorted translation and axiomatization for conditional literals. Syntactically and semantically, these constructs resemble set notation in traditional mathematics. I attempted to formalize this intuition by relating conditional literals to sets of tuples satisfying the conditions in the literal. The added axioms then mandate that the (negated) predicate in the head of the conditional literal must (not) hold for every tuple of terms in the associated set. As before, this project is designed to expand the class of ASP programs for which we have a simple, non-ground semantics.

\subsection{Conditional literals as nested implications}
\label{sec:cl}
The many-sorted approach to defining conditional literal semantics (Section~\ref{sec:ms:cl}) was appealing because, on the surface, it paralleled the informal presentation of conditional literal behavior found in teaching materials and manuals. For example, conditional literals are traditionally described (in an example-driven way) by a translation to ground forms of basic rules~\cite{potasscoGuide}. Our hope was that this new characterization could be useful from a pedagogical perspective in addition to aiding proofs of correctness. However, developing formal justifications for the equivalence of these semantics to those defined via infinitary propositional logic \cite{harliffang14} was cumbersome, and the axiomatic characterization itself was confusing. Therefore, we replaced this approach with a simpler translation to unsorted first-order logic following the intuition that conditional literals behave as nested implications. We proved the equivalence of our semantics (defined by the SM operator applied to our translated programs) to those implemented by \clingo via their equivalence to the infinitary propositional logic semantics. Once again, for tight programs this provides a first-order treatment that could be used to extend \anthem.

\subsection{Modular proofs of correctness with aggregate constraints}
Cabalar, Fandinno, and Lierler (2020) proposed a modular methodology for arguing the correctness of ASP programs. In this approach, the program is divided into various independent modules, whose behavior is captured via the SM operator. They showcase their approach using an encoding that solves the Hamiltonian Cycle problem. We extend this encoding to the Traveling Salesman problem with the addition of a constraint on the cumulative weight of the selected cycle. Importantly, this constraint uses the \texttt{sum} aggregate, which necessitated the application of our many-sorted semantics for aggregates (Section~\ref{sec:agg}). We ``recycle" the proof of correctness developed for the Hamiltonian Cycle encoding, and extend it with our own proof of correctness for the aggregate constraint. This showcases the utility of both the many-sorted aggregate semantics and the modular proof methodology. We show that our approach is also applicable to programs containing choice rules with cardinality bounds by formally proving the correctness of a Graph Coloring encoding. 
\label{sec:tsp}

\section{Ongoing Directions}

\subsection{Extending \anthem with sorts}
\label{sec:anthem}
The projects detailed in Sections~\ref{sec:agg},~\ref{sec:ms:cl}, and~\ref{sec:cl} are pieces of the foundation for a many-sorted implementation of \anthem. Developing the theory and tools for this system will likely be the basis of my dissertation, and will comprise the majority of my research activity going forward. This project presents a number of opportunities and challenges. First and foremost is the issue of automatically verifying tight programs with aggregates. \anthem requires an input specification written in first-order logic against which to compare the logic program, and currently there is no clear way to represent aggregates in this specification language. Developing such a representation and demonstrating its validity is an ongoing project for Dr. Lierler, Dr. Fandinno, and I. An alternative approach could be to enhance \anthem with the ability to check the equivalence of two programs. Then a simple, human-verified program with aggregates could act as the ``specification" and an alternative encoding (perhaps re-written for performance instead of readability) could be checked against it.

Extending \anthem to programs with aggregates will also require us to add more sort information to \anthem. Specifically, we will need to characterize the behavior of sets and set membership. The semantics introduced in Section~\ref{sec:agg} make assumptions that, for example, set membership behaves ``as expected," however such an assumption cannot be enforced in the current versions of \anthem and \vampire. The Thousands of Problems for Theorem Provers (TPTP) project has numerous partial axiomatizations of theories written in typed first-order formulas, which makes them compatible with \vampire~\cite{Sut17}. We may be able to use these as a starting point for implementing our assumptions. The theory of program completion will have to be defined for programs with multiple sorts, and the capabilities of \vampire will need to be extended as described before we can add aggregates to \anthem.

\subsection{ASP modules}
A project that I would like to undertake is the establishment of a repository of verified ASP sub-programs (``modules") that provide efficient, correct implementations of commonly encountered sub-problems. A simple example of this would be the transitive closure of a binary relation. A user who is not familiar with the nuances of ASP grounding (for instance, a civil engineer trying to apply ASP to a domain-specific problem) might write the rules
\begin{align*}
    reachable(City1,City2) &\ruleo road(City1,City2).\\
    reachable(City1,City3) &\ruleo reachable(City1,City2), reachable(City2,City3).
\end{align*}
This second rule would have an unnecessary negative impact on the grounding size and solving time of their program. A better alternative could be downloaded from an ASP repository:
\begin{align*}
    transitive(X,Y) &\ruleo edge(X,Y).\\
    transitive(X,Z) &\ruleo transitive(X,Y), edge(Y,Z).
\end{align*}
The programmer would have to define the interfaces (e.g. $road/2$ should be taken as the input relation mapped onto $edge/2$, and $reachable/2$ should be the output obtained from $transitive/2$) but otherwise such a module would be easy to integrate. For more complex sub-problems, a formal guarantee about the correctness of the module could be useful. The format of such guarantees may draw inspiration from Hoare triples~\cite{hoare1969}. For example, in this case the precondition would be the existence of a binary relation (which will be renamed $edge/2$ within the module) and the postcondition would be the existence of a binary relation representing the transitive closure of the input relation. 

Widespread community adoption of such a project would be beneficial to both developers and researchers. Reusing efficient, trustworthy code will mean more efficient, trustworthy applications of ASP in addition to reducing the time and effort required to develop them. Furthermore, creating formal guarantees of program correctness is a laborious process. Taking a modular approach where previous contributions can be shared and reused decreases this workload and might make researchers more apt to contribute such proofs. This is the strategy we undertook in our extension of the Hamiltonian Cycle problem to the Traveling Salesman problem (Section~\ref{sec:tsp}).



\section{Conclusions}
I believe that the development of trustworthy software is an important and rewarding direction of research. The formal verification of software is especially crucial in high-consequence or safety-critical systems when we need high levels of assurance. ASP has many strengths to recommend it for use in such systems.
In fact, it was the clarity and ``verifiability" of ASP programs that attracted me to this paradigm in the first place. As such, I hope that my dissertation research can expand the tools and strategies available to ASP programmers as they develop reliable software.

\bibliographystyle{eptcs}
\bibliography{bib}
\end{document}